\documentclass[preprint,12pt]{elsarticle}

\usepackage{amssymb}
\usepackage{float}
\usepackage{xspace}
\usepackage{multirow}
\usepackage{hyperref}
\hypersetup{
    colorlinks,
    linkcolor={red!50!black},
    citecolor={blue!50!black},
    urlcolor={blue!80!black}
}

\usepackage{amsmath}

\newcommand{\ttZ}{\ensuremath{\mathrm{t\overline{t}Z}}\xspace}
\newcommand{\pt}{\ensuremath{\mathrm{p_T}\xspace}}
\newcommand{\ttH}{\ensuremath{\mathrm{t\overline{t}H}}\xspace}
\newcommand{\fourtop}{\ensuremath{\mathrm{t\overline{t}t\overline{t}}}\xspace}

\journal{Nuclear Physics B}

\begin{document}

\begin{frontmatter}



\title{Differential measurements of $\ttZ$ and $\fourtop$ at large $Q^2$ at FCC-hh}


\author[label1,label2]{Louise Beriet } 
\author[label2]{Matteo Defranchis } 
\author[label2]{Birgit Stapf}
\author[label2]{Michele Selvaggi }
\affiliation[label1]{organization={Lund University},
            city={Lund},
            country={Sweden}}
\affiliation[label2]{organization={CERN},
            city={Geneva},
            country={Switzerland}}

\begin{abstract}
This contribution presents studies of differential top-quark measurements at the Future Circular Collider in its proton-proton stage (FCC-hh). The analyses target the leptonic final states of the $\ttZ$ and $\fourtop$ production processes. Particular focus is given to the high-$Q^2$ regime, where sensitivity to new physics, as encoded in an Effective Field Theory framework, is enhanced. The dataset corresponds to an integrated luminosity of 30 $\text{ab}^{-1}$ at $\sqrt{s} = 84$ TeV. The reconstructed transverse momentum $\text{p}_\text{T}(Z_{\ell\ell})$ distribution is shown to be measurable up to approximately 2 TeV with a precision of 20\% in the high energy region ($\text{p}_\text{T}(Z_{\ell\ell})$ >1.8 TeV). The scalar transverse momentum $\text{H}_\text{T}$ of the four-top production reaches 3.5 TeV with a precision of 35\%. Additionally, a dedicated study of the lepton reconstruction efficiency shows that redefining the isolation variable to account for the highly boosted objects at FCC-hh improves the total signal yield in $\ttZ$ by a factor of 1.5.
\end{abstract}


\end{frontmatter}

\section{Introduction}
\label{sec1}
Direct measurements of top quark interactions are a key focus of the Large Hadron Collider (LHC) physics program. Yet, many aspects of the top sector remain only weakly constrained. The Future Circular Collider (FCC) offers unique opportunities to expand these constraints across its two operational phases: an electron-positron collider (FCC-ee) and, later on, a proton-proton collider (FCC-hh)~\cite{FCC:2025lpp}. At FCC-ee, top couplings can be constrained through high-precision measurements near production threshold~\cite{Janot:2015yza}, while FCC-hh enables these interactions to be probed at much larger momentum transfers~\cite{Mangano:2016jyj}. \\
The FCC-hh is designed to collide protons at a current baseline scenario of $\sqrt{s} = $ 84 TeV.  In this scenario, the dataset collected by two general-purpose experiments will correspond to a total integrated luminosity of $30 \text{ ab}^{-1}$.\\
At FCC-hh energies, top quarks can be produced with transverse momenta in the multi-TeV regime, resulting in highly collimated decay products. In such configurations, leptons from the $W\to \ell \nu$ decays may lie close together, making standard isolation definition suboptimal, potentially rejecting genuine prompt leptons. 
In this context, the focus of this paper is the precision potential and experimental challenges of the FCC-hh at very large momentum transfer $Q^2$ in $\fourtop$ and $\ttZ$ production. Those measurements offer a view of the physics reach of FCC-hh in the top quark sector, complementary to direct searches for new physics. Kinematic regions with large $Q^2$ are particularly interesting within the Effective Field Theory framework, where the contribution of higher-dimensional operators typically grows with energy.\\
We report the differential yields together with estimates of the associated systematic uncertainties of such processes, as well as developments towards the improvement of the isolation criterion of leptons in the highly boosted environment of FCC-hh. The precision estimates presented in this work include statistical and detector related systematics uncertainties only. The goal of this study is to evaluate the experimental reach of FCC-hh differential measurements, and highlight the associated reconstruction challenges in a boosted environment. A realistic assessment of theoretical uncertainties, such as scale and PDF variations at FCC-hh energies, is beyond the scope of this work.

\section{Differential Analyses}
The analyses use Monte Carlo generated events at leading order (LO), using \textsc{MadGraph\_\allowbreak aMCatNLO}~\cite{Alwall:2014hca, Frixione:2021zdp} and  \textsc{Pythia8}~\cite{Sjostrand:2014zea}, processed through the fast detector simulation package \textsc{Delphes}~\cite{delphes}. The detector model is the baseline detector concept used for the FCC-hh physics studies, which has been described in Ref.~\cite{FCC-hhCDR, Mangano:2022ukr} and its  \textsc{Delphes} parameterization is described in Ref.~\cite{Selvaggi:2717698}.

\subsection{$\fourtop$}
\begin{figure}[htb]
    \centering
    \includegraphics[width=0.45\linewidth]{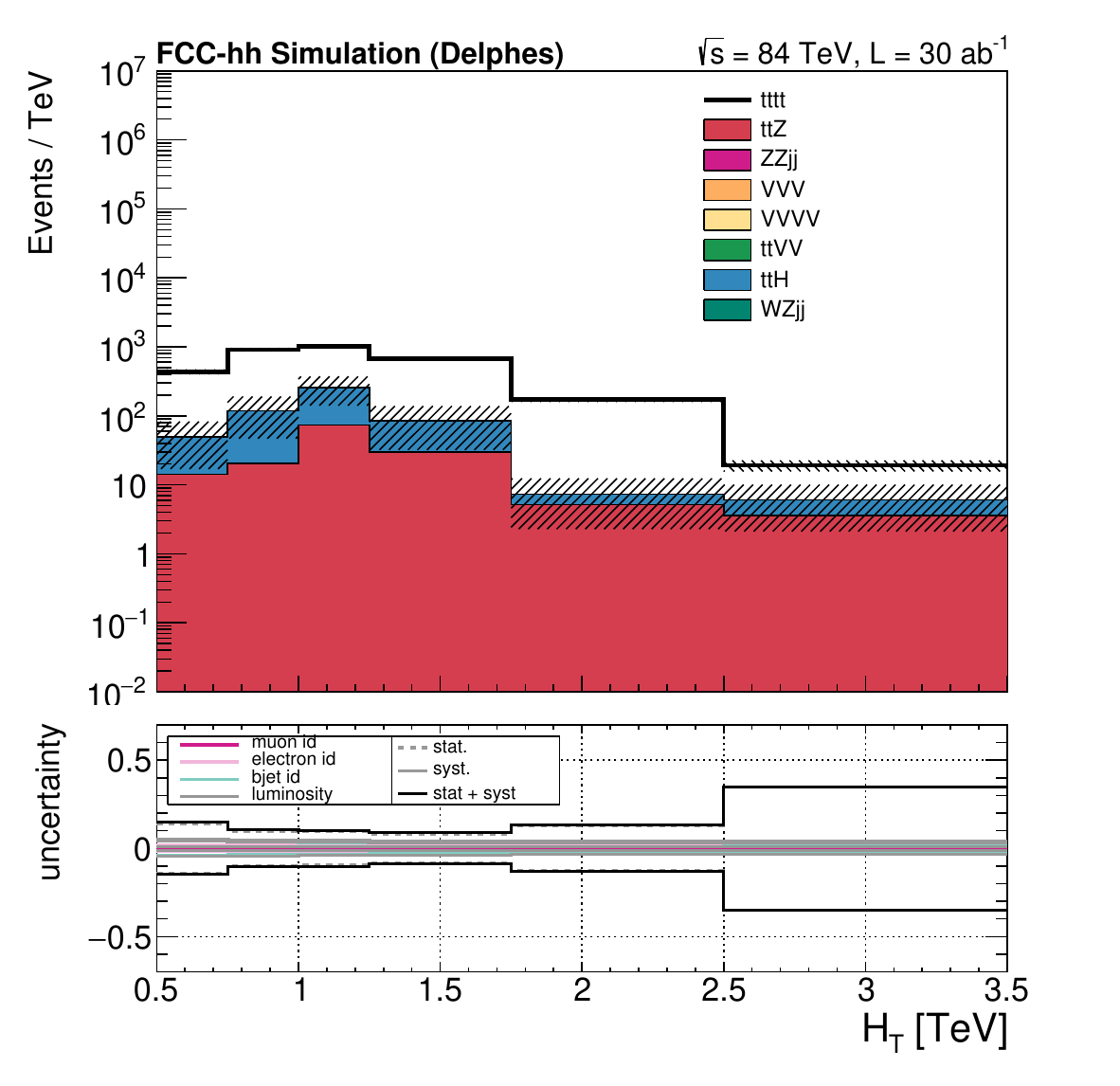}
    \includegraphics[width=0.5\linewidth]{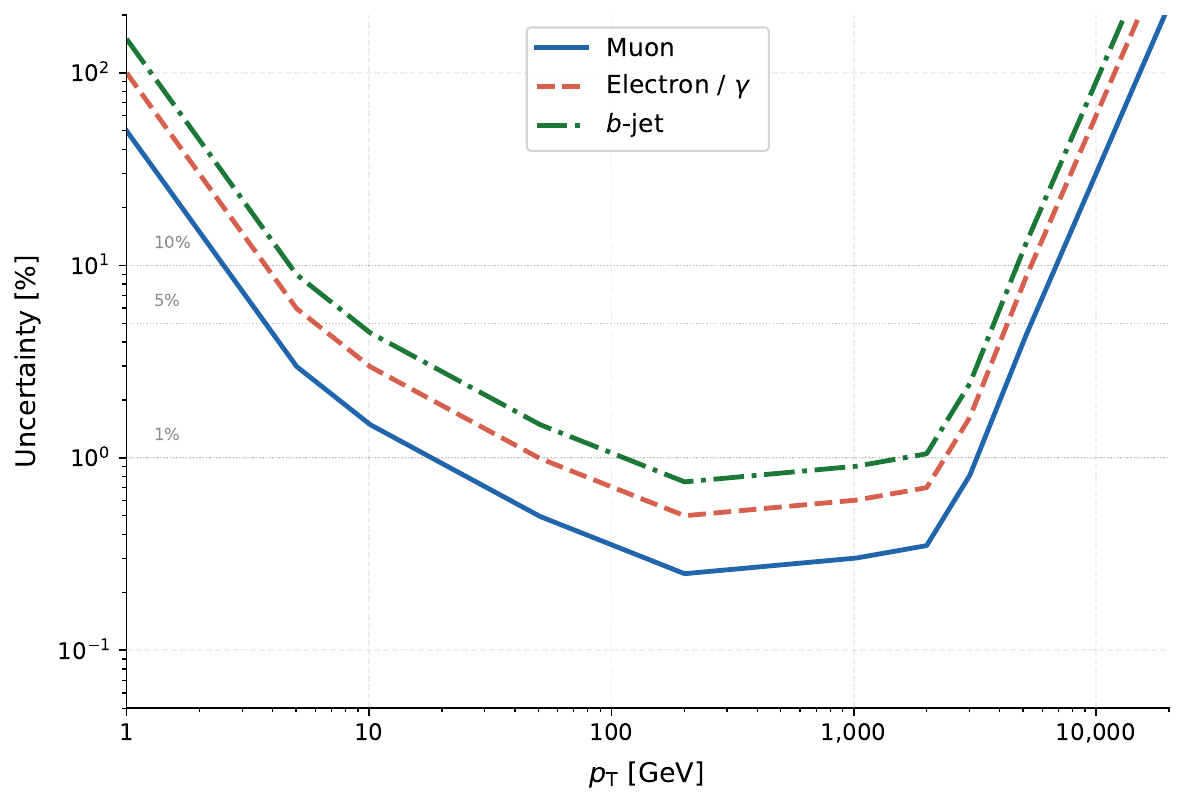}
    \caption{\textit{Left}: Distribution of the reconstructed $H_T(\fourtop)$ after event selection. Bottom panel displays the relative contributions from statistical and systematic uncertainties. \textit{Right:} Uncertainty on the electron and muon identification efficiencies and b-jet tagging efficiency as a function of $p_T$.  }
    \label{fig:4t}
\end{figure}

The four-top quark production is a rare process at the LHC, with a theoretical cross-section at $\sqrt{s}= 13.6$ TeV of 12.3 fb (NLO+NLL')\cite{Beekveld:2025}. FCC-hh will significantly increase the available statistics for it, enhancing the cross-section to 1.6 pb (LO), and the integrated luminosity by about $\times 5$ compared to HL-LHC. The final state considered for this study is composed of four leptons, two of which are positively charged and two negatively charged, at least three b-tagged jets (b-jets), and missing energy from the neutrinos. It follows the leptonic decays of the four tops:  $t \rightarrow W^+b \rightarrow(\ell^+\nu)b \text{ or }\bar{t}\rightarrow W^-\bar{b}\rightarrow(\ell^-\bar{\nu})\bar{b}$. The background considered is made of all processes producing a similar final state signature, namely: VVV, VVVV, $\ttZ$, $\bar{\text{t}}$tVV, $\ttH$, ZZ+2 jets, where V is a vector boson. The topic of experimental separation of $\fourtop$ and $\text{t}\bar{\text{t}}\text{tW}$ is an active area of research, and is not addressed in this study. A strict selection is applied to all events, as described in \autoref{tab:cutflow}. 
\renewcommand{\arraystretch}{1.3}
\begin{table}[htb]
\begin{tabular}{|l|p{4cm}|p{5cm}|}
\hline
                         & \multicolumn{1}{c|}{$\ttZ$}                                    & \hspace{2cm}$\fourtop$                \\ \hline
Acceptance               & \multicolumn{2}{c|}{$p_T >30$ GeV , $|\eta|$ \textless 4}                                  \\ \hline
\multirow{3}{*}{Leptons} & \multicolumn{2}{c|}{4 isolated leptons}                                                    \\ \cline{2-3} 
                         & \multicolumn{1}{l|}{$e^\pm e^\mp(\text{or }\mu^\pm\mu^\mp)$ and $e\mu$}           & $e^\pm e^\pm \mu^\mp \mu^\mp$ \\ \cline{2-3} 
                         & \multicolumn{1}{l|}{$|m_Z - m_{e^+e^-(\mu^+\mu^-)}| < 10$ GeV} &                           \\ \hline
b-tagged jets            & \multicolumn{1}{l|}{$N_{b-jets}$ = \{1,2\}}                    & $N_{b-jets} > 2$          \\ \hline
\end{tabular}
\caption{Event selection for the $\ttZ$ and $\fourtop$ measurements.}\label{tab:cutflow}
\end{table}\\
Although the four-top final state contains four b quarks at tree level, the analysis requires $N_{b-jets} >2$ rather than 4 identified b-jets to preserve signal efficiency with finite b-tagging capability. Processes such as $t\bar{t}WW$, which contain at most two b quarks from the top decays, are strongly suppressed by the requirement $N_{b-jets}>2$. The lepton flavor and sign requirements are particularly efficient in suppressing processes including Z production, as a final state $e^\pm e^\pm \mu^\mp \mu^\mp$ cannot originate from Z decays. Most of the reducible background is suppressed after this requirement. \\
The chosen variable correlated with the momentum transfer in the process is the transverse momenta scalar sum $\text{H}_\text{T}  = \sum \vec{p_T}(\ell)+\vec{p_T}(jets)$.\\
The $\text{H}_\text{T}$ distribution after selection is shown in \autoref{fig:4t}, left. The quoted uncertainties include statistical contribution, a 1\% luminosity uncertainty, and detector related uncertainties associated with lepton and b-jets identification efficiencies. The latter are parametrized as functions of the transverse momentum of the reconstructed objects (see \autoref{fig:4t}, right), and originate from the finite statistics of control high $p_T$ Drell-Yan samples at the FCC-hh.  A 35\% precision in the highest bin (2.5 to 3.5
TeV), and a global 3.4\% precision on the $\fourtop$ production can be achieved in its fully leptonic final state.
\subsection{$\ttZ$}
\begin{figure}[htb]
    \centering
    \includegraphics[width=.45\linewidth]{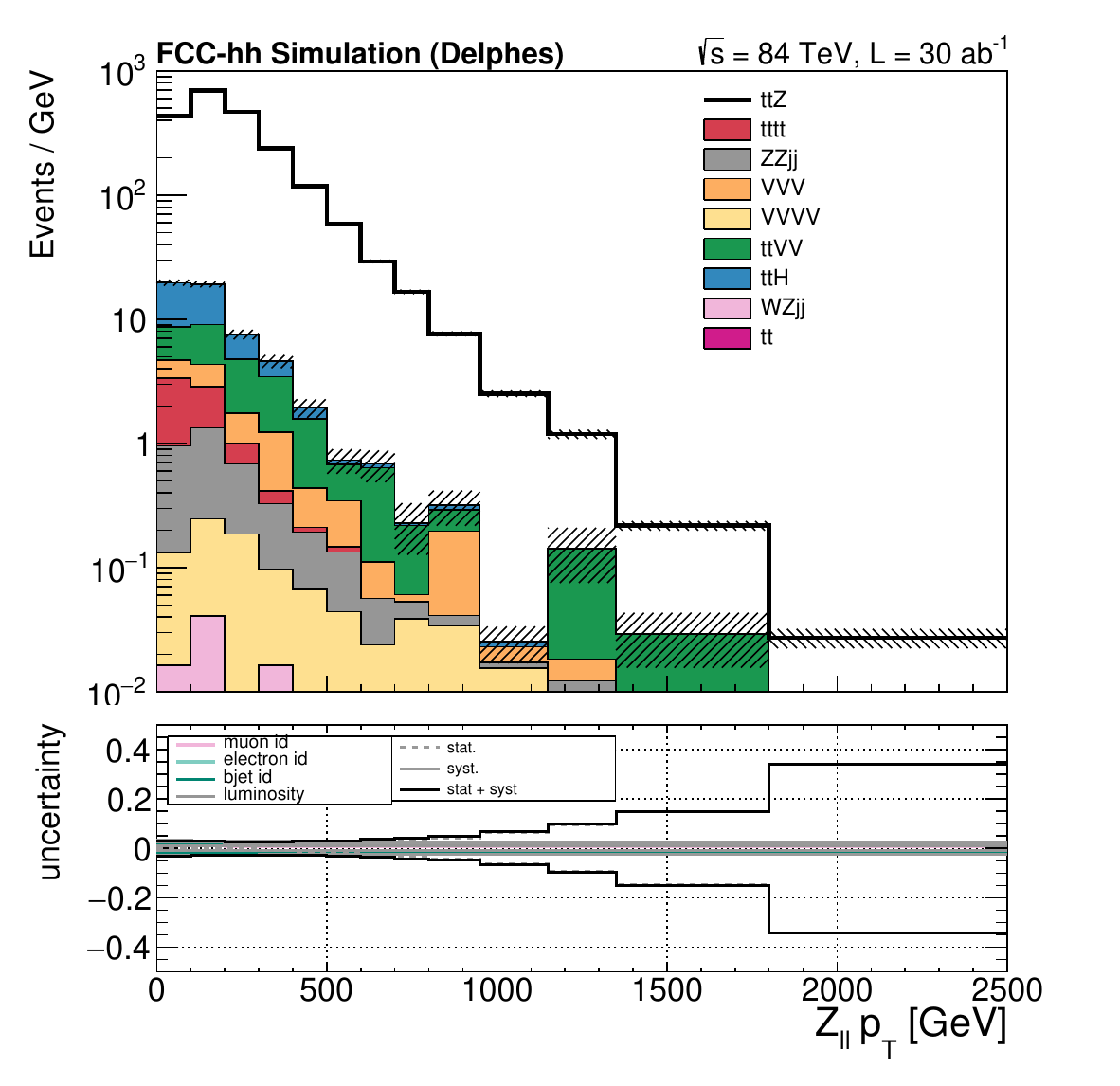}
    \includegraphics[width =.45\linewidth]{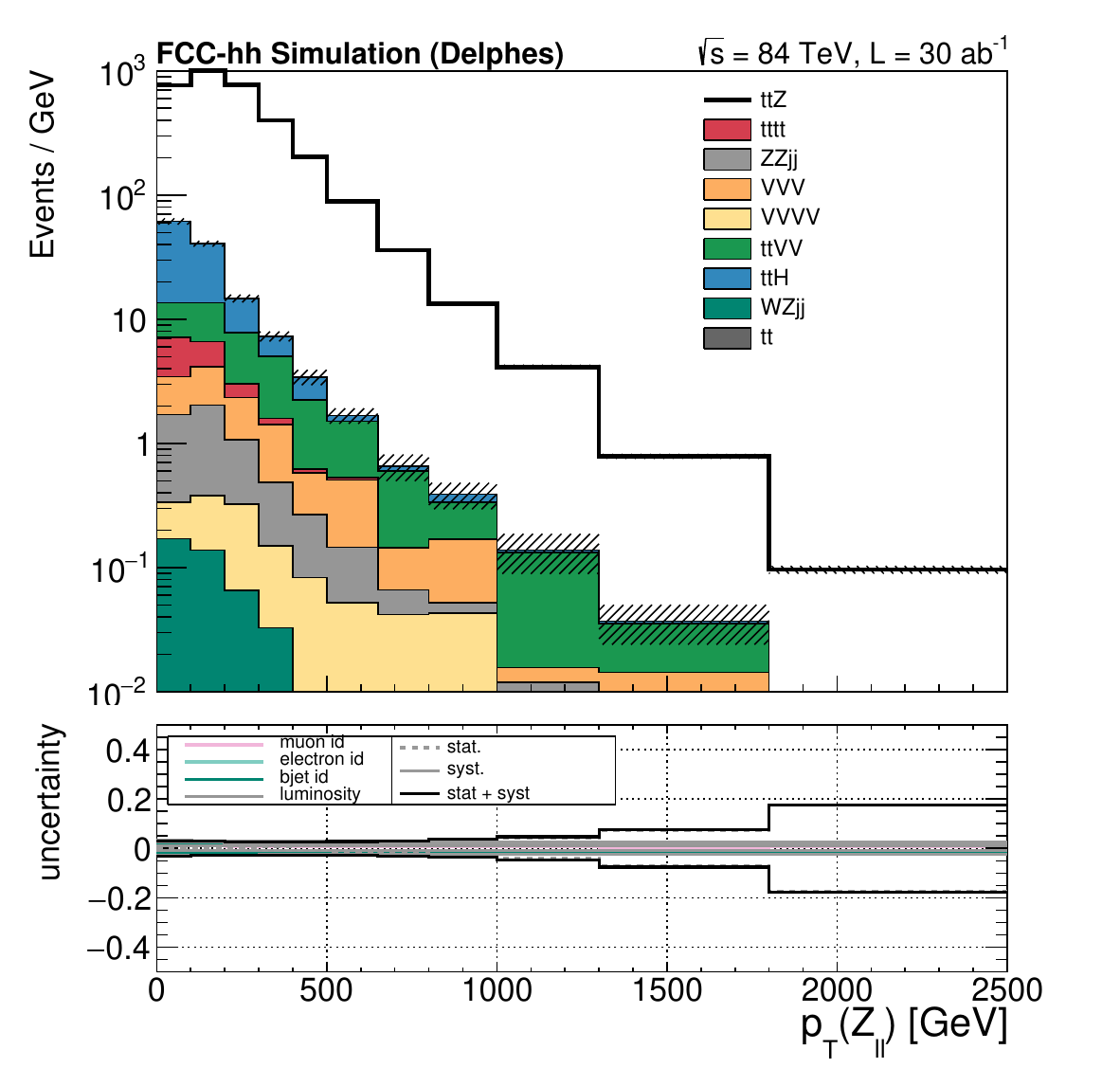}
    \caption{$\text{p}_\text{T}(Z_{\ell\ell})$ distribution of the $\ttZ$ process after event selection, before (left) and after (right) isolation optimization. Bottom panels display the relative contributions of statistical and systematic uncertainties.}
    \label{fig:ttZ}
\end{figure}

The associated production of a top quark pair and a Z boson ($\ttZ$) is a key process to probe the coupling of the top quark to the electroweak sector. At $\sqrt{s} =84$ TeV, the cross-section is $\sigma_{\ttZ} = 28.2$ pb (LO), 32 times greater than the cross-section obtained during the LHC Run 2, at $\sqrt{s} = 13$ TeV. We focus on the decay channel where the Z boson and top quark pair decay leptonically, for its high purity. The final state consists of two leptons opposite in charge and with the same flavour coming from the Z, two b-jets, missing energy, and two leptons opposite in charge coming from the tops. The following processes are considered as background: $\fourtop$, $\ttH$, VVV, VVVV, WZ+2 jets, ZZ+2 jets, $\text{t}\bar{\text{t}}$VV, and t$\bar{\text{t}}$. The event selection is described in \autoref{tab:cutflow}. In events containing more than one possible opposite-sign same-flavour lepton pair, the Z candidate is defined as the pair whose invariant mass is closest to the nominal Z-boson mass: 91.2 GeV \cite{ParticleDataGroup:2024cfk} .\\
The transverse momentum of the lepton pair originating from the Z is defined as the vector sum: $\vec{\mathrm{p_T}}(Z_{\ell\ell})= \vec{\mathrm{p_T}}(\ell_1) +\vec{\mathrm{p_T}}(\ell_2)$. With $\ell_1 = \mu^\pm \text{ or } e^\pm$ and $\ell_2 = \mu^\mp \text{ or }e^\mp$. The $\mathrm{p_T}(Z_{\ell\ell})$ distribution after selection (without, left, or with, right, improved isolation) is shown in \autoref{fig:ttZ}. A precision of $\approx 35$\% can be achieved in the high-$p_T$ kinematic region ($p_T(Z_{\ell\ell}) >1.8$ TeV), and an expected global precision of 2.6\% on the $\ttZ$ production cross section in the fully leptonic channel. 

\section{Lepton Isolation}
Reconstructed leptons are classified into prompt and non-prompt based on the isolation variable defined as the ratio of the sum of all transverse momenta ($p_T$) of the objects ($p_i$) around that lepton within a cone $\Delta R$, over the transverse momentum of that lepton ($p_T(\ell)$):
\begin{equation}\label{eq:1}
     \text{Iso} = \frac{\Sigma_i^{p_{Tmin}>y\hspace{3pt}; \hspace{3pt}\Delta R < x } p_T(p_i)}{p_T(\ell)}
\end{equation}

The results presented in \autoref{sec1} were obtained using an isolation value of 0.1 for muons, 0.2 for electrons, for a cone size of $\Delta R$ = 0.2, used by default in other FCC-hh studies. Photons radiated by electrons can produce additional nearby energy deposits, while muons behave approximately as minimum ionizing particles. Slightly looser isolation requirements are typically applied to electrons to account for this difference in detector signature.\\
The definition in \autoref{eq:1} works well under LHC-like conditions. However, for FCC-hh, operating at a much higher center-of-mass energy, the environment will feature much more boosted objects. High momentum decay products of heavy particle will tend to be more collimated. In the $\ttZ$ case, very close leptons from boosted Z decay can end up overlapping within their $\Delta R$ cone radius, artificially excluding each other from passing the isolation cut.\\
To overcome this issue, we redefined the isolation excluding leptonic activity from the $\Delta R$. \\
The optimal isolation threshold value is computed based on the $\ttZ$ sample. The True Positive Rate (TPR, or efficiency) and False Positive Rate (FPR, or inefficiency) are computed for different values of the isolation and for different $\Delta R$. The optimal threshold value is obtained by optimizing the Youden index: J = TPR $-$ FPR. The new isolation points $\text{Iso}(p_{Tmin} = 0.5 \text{GeV}, \Delta R = 0.1) = $ 0.181 and 0.235, for muons and electrons respectively, improve the prompt-lepton efficiency (TPR) from 73\% to 94\% for electrons, and 84\% to 96\% for muons. However, the FPR is increased from 3\% to 8\% for electrons, and from 1\% to 3\% for muons. Given the high purity of the channel, recovering additional prompt leptons is more beneficial than a small rise in non-prompt contamination.   \\
With the optimized isolation,  the new $p_T(Z_{\ell\ell})$ distribution shows significant improvement. Precision in the last bin is improved by a factor of two, with the total uncertainty decreasing from 38\% to 19\%. The improvement is driven by the increased signal statistics.

\section{Conclusion}
The expected precision of two differential measurements, $\ttZ$ and $\fourtop$, in the top quark sector at FCC-hh has been demonstrated. The measurements specifically aim to probe the high $Q^2$ region, where sensitivity to SM deviations encoded in Effective Field Theories is higher. The large dataset of FCC-hh allows to focus on rare multi-leptons final states, where background contamination is reduced. 
The results obtained show the remarkable reach of FCC-hh: the $\ttZ$ differential distribution extends to $\text{p}_\text{T}(Z_{\ell\ell}) =$ 2.5 TeV with about 20\% precision in the high-$\pt$ region, while the $\fourtop$ distribution reaches $\text{H}_\text{T}$ = 3.5 TeV with 35\% precision.
In addition, the study of lepton isolation in highly boosted topologies shows that standard isolation definitions become suboptimal in the FCC-hh environment. By redefining the isolation variable to exclude nearby leptonic activity and re-optimizing the threshold value, the prompt-lepton efficiency increases to about 95\%. This improvement enhances the signal yield and leads to a factor of two gain in precision in the last bin of the $\pt(Z_{\ell\ell})$ distribution.
 \bibliographystyle{elsarticle-num} 
 \bibliography{SummerCERNFCChh}






\end{document}